# Analysis of Probabilistic multi-scale fractional order fusion-based de-hazing algorithm

U. A. Nnolim
*Department of Electronic Engineering*
*University of Nigeria, Nsukka, Enugu*

**Abstract**

*In this report, a de-hazing algorithm based on probability and multi-scale fractional order-based fusion is proposed. The proposed scheme improves on a previously implemented multiscale fraction order-based fusion by augmenting its local contrast and edge sharpening features. It also brightens de-hazed images, while avoiding sky region over-enhancement. The results of the proposed algorithm are analyzed and compared with existing methods from the literature and indicate better performance in most cases.*

**Introduction**

The end product of image de-hazing includes contrast maximization and improved visibility of details. Single image de-hazing has become an attractive option over multi image methods. This is primarily due to feasibility and convenience since single image-based methods do not require multiple instances of the same image scene [1]. There are four categories of image de-hazing namely enhancement, restoration, fusion and deep learning-based approaches [2]. These categories range from simple to complex architectures based on works from the literature [2]. The fusion and deep learning-based methods are relatively recent, while incorporating various techniques. However, most algorithms work best for either single image de-hazing or underwater image enhancement. This is notwithstanding the similarities between hazy and underwater images, which both suffer from poor visibility due to inadequate contrast. The deep learning-based approaches are highly involved, requiring a large image database for training. Furthermore, there is need for graphic processing units (GPUs) to reduce the execution time of such methods. Other authors utilize other methods to accelerate and optimize their algorithms [3]. These all add to the problems of colour distortion, minimal contrast enhancement, poor colour rendition and minimal colour correction for several images. Furthermore, a number of restoration and fusion-based methods require the manual setting of parameters, which may not yield the best result for all images.

**2. Background and motivation**

In previous work, we developed a versatile multi-scale fractional order fusion-based de-hazing algorithm, which could effectively process both underwater and hazy images without modification. Furthermore, the algorithm was completely automated, with all vital parameters computed from the input image. Results showed promising results and fast operation and minimal runtime, which surpassed several of the existing algorithms. The algorithm utilized fractional-based filters at varying scales to process the image, while fusing results together to generate a much more detailed image. However, this proposed algorithm yielded minimal local contrast enhancement in addition to edge over-enhancement for certain images with a high amount of detail. Furthermore, the algorithm yielded dark regions for some processed images. Thus, in this work, we combined tonal correction and localized contrast operators with a revised formulation for the fractional multi-scale-based algorithm. This led to the improvement and amplification of local contrast enhancement, while brightening de-hazed images, avoiding colour distortion, sky region and edge over-enhancement.

## 3. Proposed algorithm

The proposed scheme utilizes probabilistic and simultaneous estimation of illumination and reflectance components developed by Fu et al [4]. This is combined with the bilateral filter by Tomasi and Manduchi [5] for multi-scale illumination estimation rather than the Gaussian filter. This is in addition to the revised fractional multi-scale fusion-based algorithm used in improving the fine detail at each stage. This enables the localized contrast enhancement and avoidance of edge over-sharpening due to the non-linear means (bilateral) filter-based estimation. The system block diagram of the proposed scheme is shown in Fig. 1. The full details of the proposed scheme and its variants can be found in [6].

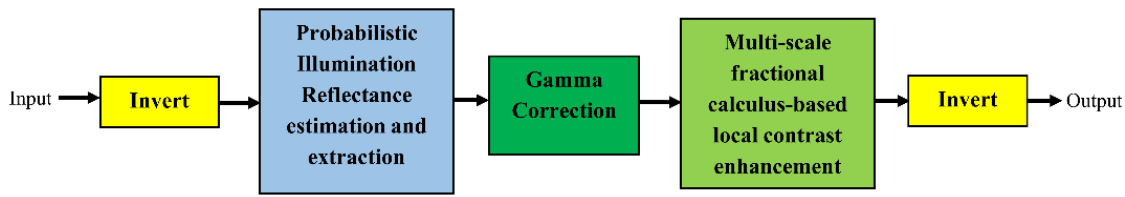

Fig. 1. Probabilistic illumination correction + fractional-derivative-based multi-scale de-hazing system

## 4. Experiments and results

This section presents the results and comparisons of the proposed scheme with both hazy and underwater image enhancement algorithms from the literature using subjective evaluation, objective assessment using various metrics, image statistics and histogram information.

### 4.1. Underwater images

The proposed scheme was tested using numerous underwater image scenes and datasets commonly found in the literature. Furthermore, we compared results using popular and recent state-of-the-art underwater image enhancement algorithms. The result is shown in Fig. 2(a).

Fig. 2(a) consists of a figure from [7] (amended with PA-1 results), which includes the following algorithms: Contrast Limited Adaptive Histogram Equalization (CLAHE), Retinex, White balance, combined Methods of He et al, Zhu et al and Non-local de-hazing, DehazeNet, cycle-consistent adversarial networks (CycleGAN), Li's method and the adjusted CycleGAN to compute structural similarity index metric (SSIM) loss (CycleGAN+SSIM Loss size) [7]. For all tables, bolded black indicates the best values while bolded red depicts the second best results.

In Fig. 2(b)(1) and (2), we compare the PA with image results obtained with other algorithms from the literature such as Ancuti et al [8], Bazeille et al [9], Carlevaris-Bianco et al [10], Chiang et al [11], Galdran et al [12], Serikawa and Lu [13], compared with PA. The images produced by Ancuti et al, Me et al and PA are the best compared with other methods. However, observation of the RGB histogram plots show that PA yields the most stretched colour histogram, indicating highest contrast enhancement. This is also observed I Table 1, which presents the mean and standard deviation values for the hue saturation and value components of the original and processed underwater images.

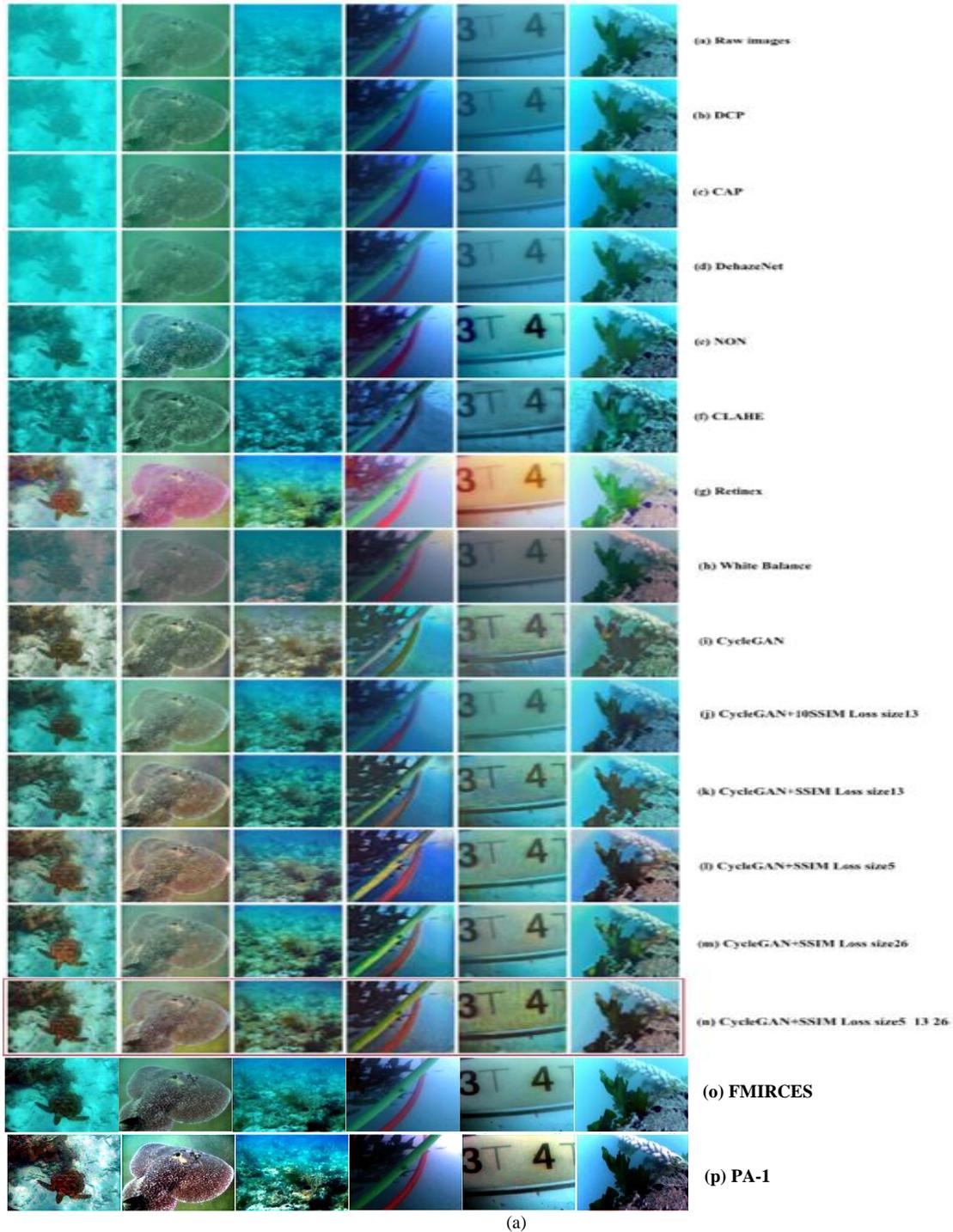

Fig. 2(a) Figure from [7] amended with visual result of FMIRCES and PA-1 for comparison

Overall, PA yields the highest standard deviation of saturation and value components, which indicate improved colour rendition and contrast enhancement respectively. This is also observed I Table 1, which presents the mean and standard deviation values for the hue saturation and value components of the original and processed underwater images. The colour cast is considerably eliminated in the image obtained with PA compared with the other algorithms. This is also observed in Fig. 2(c), which once more shows that PA surpasses several of the other algorithms by improving contrast and colour rendition, while eliminating colour cast. It yields results similar to Galdran et al but with sharper focus and the histogram would be similar to those of Fig. 2(b)(1) and (2), with highly stretched red, green and blue colour channels or highly enhanced saturation and value components.

We also utilize the mean (mu), central moment of order 2 (mu_2) [14], standard deviation (sigma), skewness (gamma) [14], momental skewness (alpha) [14] and kurtosis (kappa) [14] to measure the various properties of the processed images. The mean indicates the centeredness of the histogram of the processed image. The standard deviation is linked to contrast, while the skewness determines the directional bias of the image histogram and the kurtosis defines the narrowness of the peak of the histogram. Thus, a more stretched out histogram would have a kurtosis value less than 3, while an unmodified histogram with a sharp peak would have a value greater than or equal to 3. Looking at the results in Table 2, the results of PA yield the lowest kurtosis values for the red, green and blue channels. This is expected since PA stretches out the histogram and we also expect the standard deviation to increase as the histogram is stretched, indicating more spread [15]. Also, this shows in the standard deviation values of the green and blue channels in Table 2, whose histograms have been considerably stretched compared to the red channel, improving the contrast. These numerical results are in line with the image histogram plots in Fig. 2(b)(3). We will also observe that the PA indicates high localized contrast for certain images, based on the histogram plots.

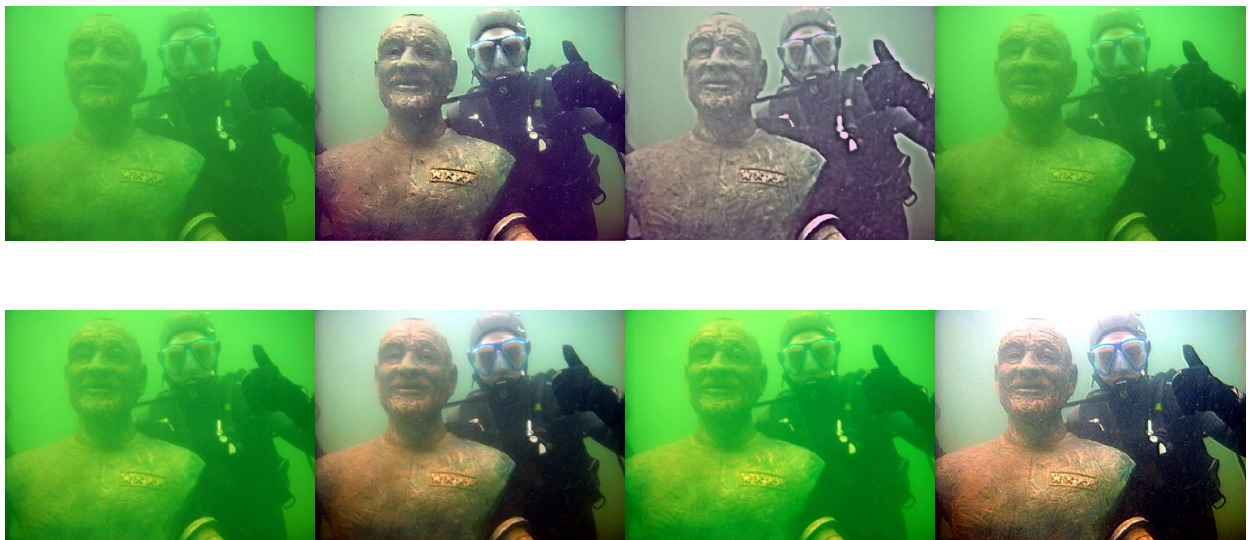

(1)

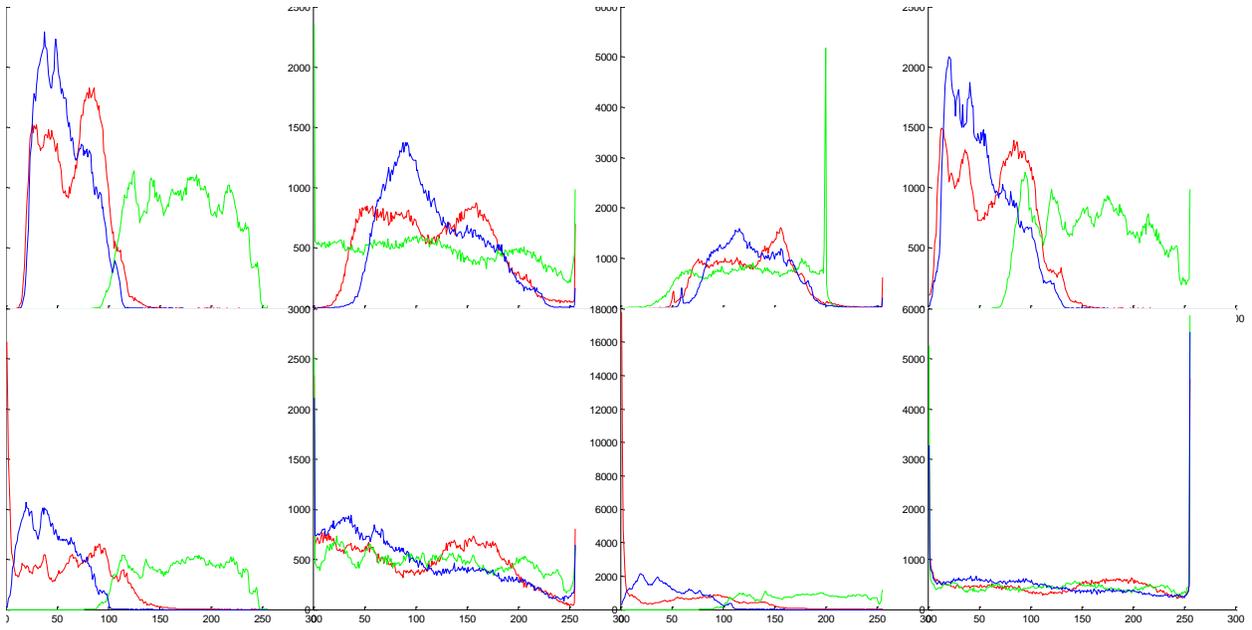

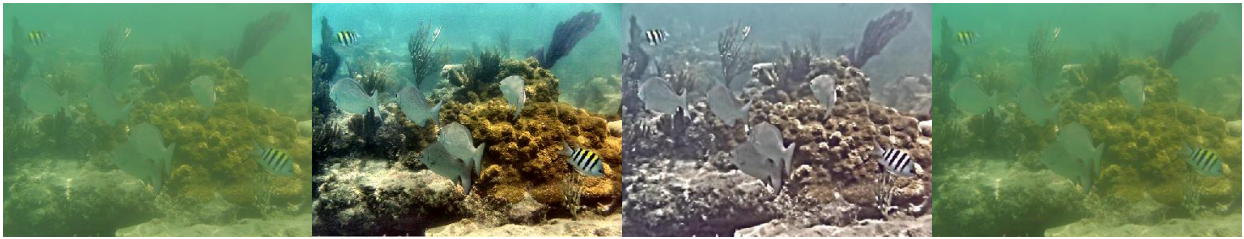

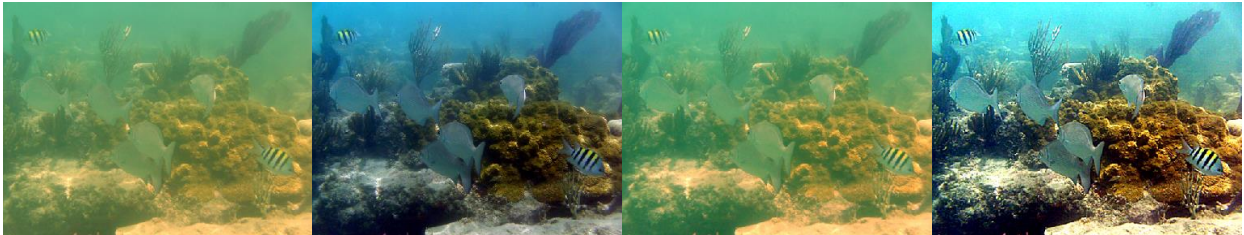

(2)

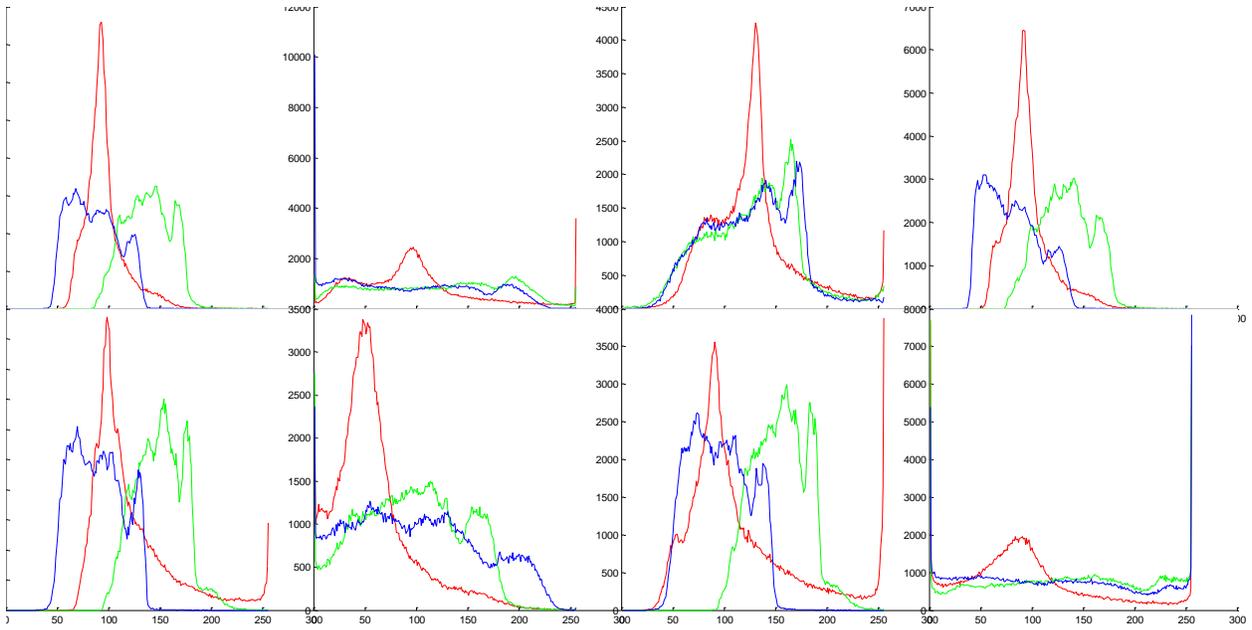

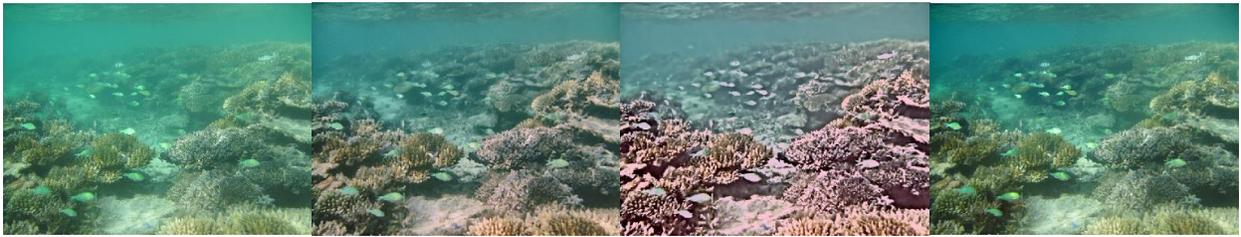

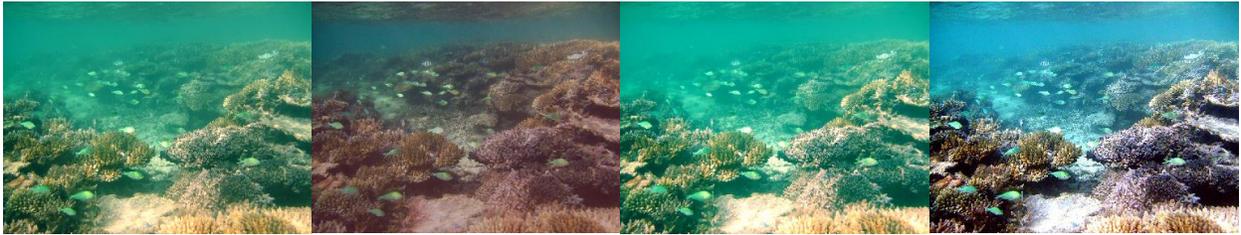

(3)

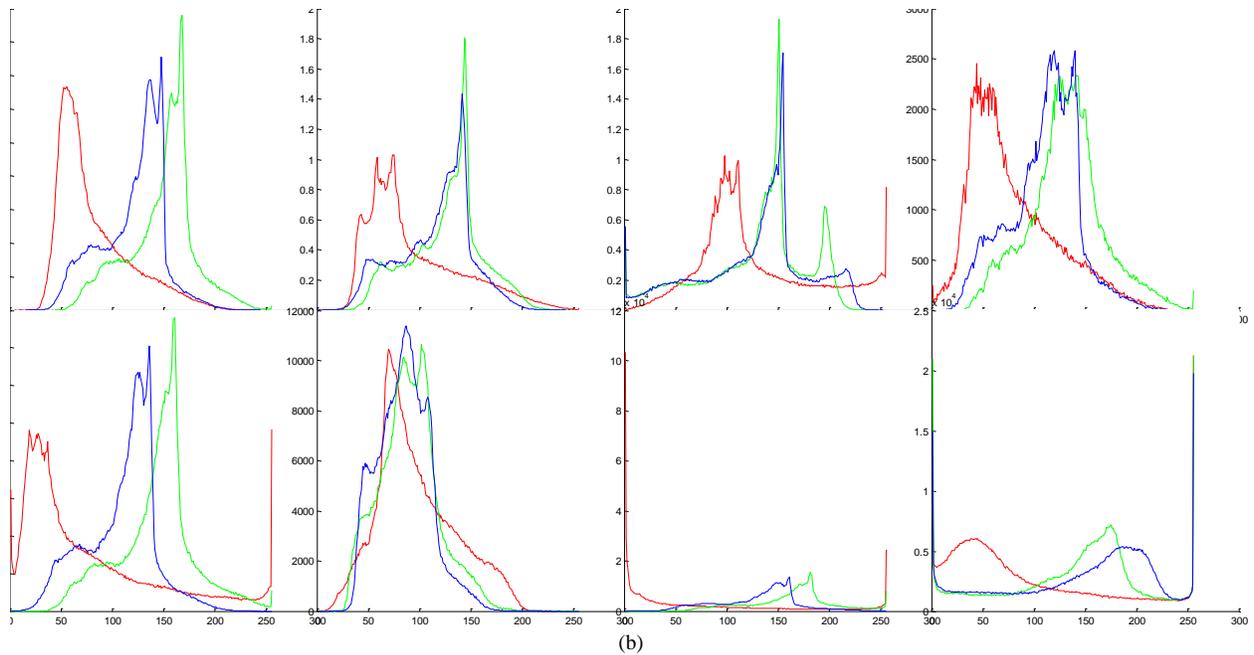

| Original image | Ancuti | Bazeille | Carlevaris-Bianco et al |
|---|---|---|---|
| Chiang & Chen | Galdran et al | Serikawa & Lu | PA |

**KEY**

Fig. 2 (b) processed images using various algorithms and corresponding histograms with key to figures

Table 1. Statistical values for hue saturation and value channels of processed images from Fig. 2(b)(1) and Fig. 2(b)(2)

| Image | *Diver and statue (Fig. 2(b)(1))* | | | | | | | |
|---|---|---|---|---|---|---|---|---|
| Algos \\ Statistical measures | Original image | PA | Ancuti et al | Bazeille et al | Carlevaris-Bianco et al | Chiang & Chen | Galdran et al | Serikawa & Lu |
| *H_mean* | 0.3232 | 0.3593 | **0.4674** | 0.4667 | 0.3197 | 0.3165 | 0.3270 | 0.3230 |
| *S_mean* | 0.7001 | 0.3933 | 0.3406 | 0.1684 | 0.7225 | 0.7965 | 0.4165 | **0.8284** |
| *V_mean* | 171.5676 | 141.5357 | 136.2584 | 138.1032 | 158.6579 | 170.7642 | 123.9213 | **177.6764** |
| *H_std* | 0.0204 | 0.2575 | **0.2791** | 0.2756 | 0.0265 | 0.0328 | 0.2029 | 0.0399 |
| *S_std* | 0.0708 | **0.2855** | 0.2306 | 0.1045 | 0.1050 | 0.1152 | 0.2498 | 0.1365 |
| *V_std* | 38.7415 | **78.5256** | 57.8897 | 40.0237 | 48.1707 | 40.2752 | 72.7430 | 44.1202 |

| Image | *Fishes (Fig. 2(b)(2))* | | | | | | | |
|---|---|---|---|---|---|---|---|---|
| Algos \\ Statistical measures | Original image | PA | Ancuti et al | Bazeille et al | Carlevaris-Bianco et al | Chiang & Chen | Galdran et al | Serikawa & Lu |
| *H_mean* | 0.3045 | 0.3886 | 0.3050 | 0.4096 | 0.2905 | 0.2584 | **0.4478** | 0.2900 |
| *S_mean* | 0.4291 | **0.5609** | 0.4830 | 0.1588 | 0.4459 | 0.4592 | 0.5559 | 0.4966 |
| *V_mean* | 137.5797 | 148.0420 | 128.3777 | 136.5561 | 132.4112 | 153.8955 | 110.3785 | **160.6456** |
| *H_std* | 0.0655 | **0.2220** | 0.1853 | 0.3026 | 0.0712 | 0.0940 | 0.1864 | 0.1098 |
| *S_std* | 0.0629 | **0.2521** | 0.2283 | 0.0954 | 0.0722 | 0.0788 | 0.2481 | 0.0851 |
| *V_std* | 22.2850 | **72.3171** | 64.8840 | 43.3192 | 25.1146 | 29.9721 | 54.9853 | 34.5472 |

Table 2. Statistical values for red, green and blue channels of processed images from Fig. 2(b)(3)

| components | r | g | b |
|---|---|---|---|
| mu | 0.31824 | 0.5787 | 0.46875 |
| mu_2 | 0.01944 | 0.01671 | 0.01478 |
| sigma | 0.13942 | 0.12929 | 0.12156 |
| gamma | 1.17344 | -0.36641 | -0.43084 |
| alpha | 0.58672 | -0.1832 | -0.21542 |
| kappa | 3.80406 | 3.20924 | 2.81875 |
| **Original image** | | | |
| mu | 0.37048 | 0.4939 | 0.45048 |
| mu_2 | 0.03291 | 0.01928 | 0.01973 |
| sigma | 0.18141 | 0.13886 | 0.14047 |
| gamma | 0.95014 | -0.25917 | -0.37492 |
| alpha | 0.47507 | -0.12958 | -0.18746 |
| kappa | 3.11475 | 2.74043 | 2.5583 |
| **Ancuti et al** | | | |
| mu | 0.49913 | 0.50073 | 0.50056 |
| mu_2 | 0.0503 | 0.03964 | 0.04233 |
| sigma | 0.22428 | 0.1991 | 0.20575 |
| gamma | 0.67522 | -0.65467 | -0.4715 |
| alpha | 0.33761 | -0.32734 | -0.23575 |
| kappa | 2.63607 | 2.78701 | 2.71791 |
| **Bazeille et al** | | | |
| mu | 0.30947 | 0.51046 | 0.427 |
| mu_2 | 0.02787 | 0.02094 | 0.01774 |
| sigma | 0.16693 | 0.1447 | 0.13321 |
| gamma | 0.86462 | 0.14203 | -0.20723 |
| alpha | 0.43231 | 0.07102 | -0.10362 |
| kappa | 3.1859 | 3.34694 | 3.06297 |
| **Carlevaris-Bianco et al** | | | |
| mu | 0.29426 | 0.55939 | 0.42587 |
| mu_2 | 0.06486 | 0.022 | 0.01748 |
| sigma | 0.25468 | 0.14833 | 0.13222 |
| gamma | 1.21131 | -0.13494 | -0.36391 |
| alpha | 0.60566 | -0.06747 | -0.18196 |
| kappa | 3.57397 | 3.37775 | 2.98694 |
| **Chiang and Chen** | | | |
| mu | 0.37515 | 0.3492 | 0.32988 |
| mu_2 | 0.02244 | 0.01281 | 0.00981 |
| sigma | 0.1498 | 0.11319 | 0.09905 |
| gamma | 0.53901 | 0.34197 | 0.32637 |
| alpha | 0.2695 | 0.17099 | 0.16319 |
| kappa | 2.71403 | 3.24983 | 3.0802 |
| **Galdran et al** | | | |
| mu | 0.27137 | 0.63599 | 0.51275 |
| mu_2 | 0.1006 | 0.0256 | 0.02332 |
| sigma | 0.31717 | 0.16001 | 0.1527 |
| gamma | 1.10219 | -0.24416 | -0.29823 |
| alpha | 0.5511 | -0.12208 | -0.14911 |
| kappa | 2.92141 | 3.0903 | 3.00608 |
| **Serikawa & Lu** | | | |
| mu | 0.34689 | 0.53313 | 0.56371 |
| mu_2 | 0.07853 | 0.06669 | 0.076 |
| sigma | 0.28024 | 0.25824 | 0.27569 |
| gamma | 0.92815 | -0.44633 | -0.56366 |
| alpha | 0.46407 | -0.22316 | -0.28183 |
| kappa | 2.78492 | 2.57673 | 2.28078 |
| **PA** | | | |

The remaining images in Fig. 2(c) to 2(e) compare the results of PA with the methods by Ancuti et al [8], Bazeille et al [9], Carlevaris-Bianco et al [10], Chiang & Chen [11], Galdran et al [12] and Serikawa & Lu [13] with their corresponding image histograms. Results indicate consistent trends for PA in terms of histogram stretching effects and once more shows the most spread out histogram.

Since underwater and hazy images share similar degradation effects, we expect the results of the hazy images to be similar to the underwater image results. Also note that PA removes colour cast without modification of the algorithm and also incorporates local contrast enhancement without histogram equalization-based methods. The histograms in general show highly stretched colour channels compared with the original images. These have been supported with computation of statistical values enumerated earlier in this section. This utilization of image colour histograms to profile various underwater images has been used in previous work [16] [17] [18] [19] with interesting outcomes and once more helps in visualization of the properties of the PA.

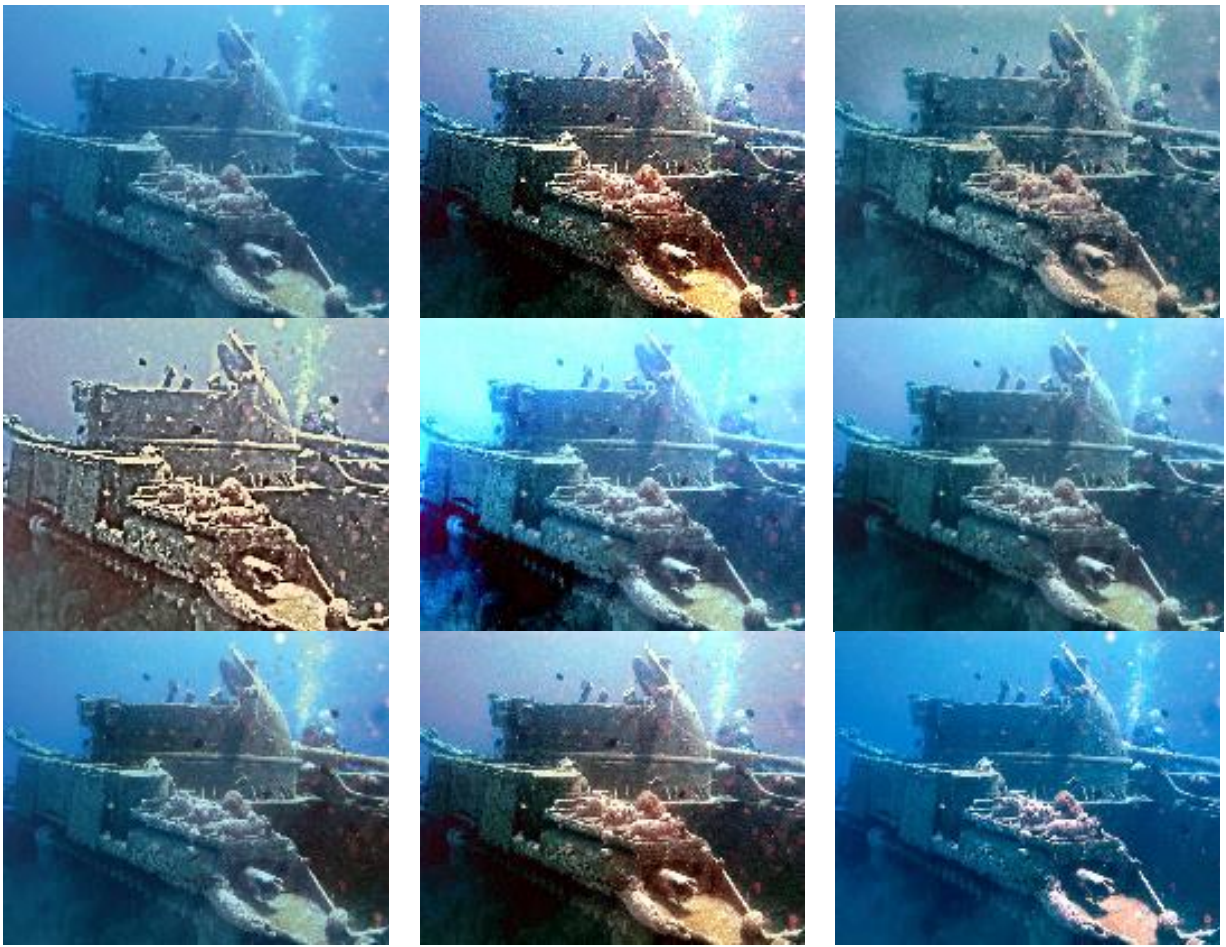

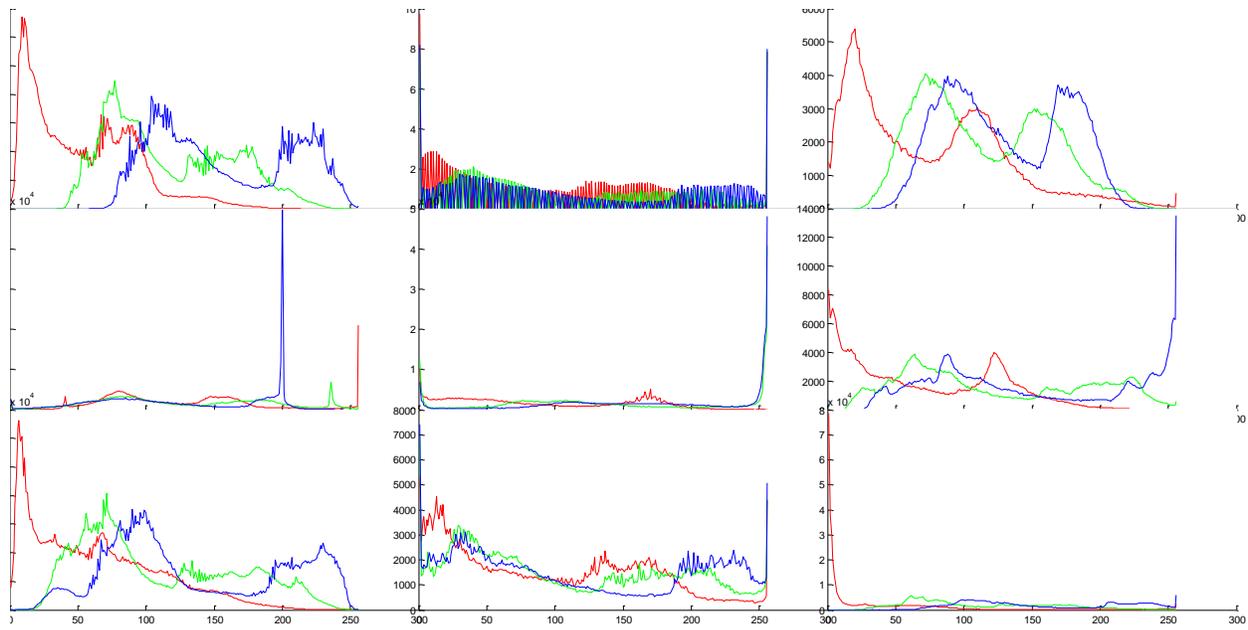

| Original image | PA | Ancuti |
| --- | --- | --- |
| Bazeille et al | Carlevaris-Bianco et al | Chiang & Chen |
| Fattal | Galdran et al | Serikawa & Lu |

KEY

Fig. 2 (c) processed images using various algorithms and key to figures

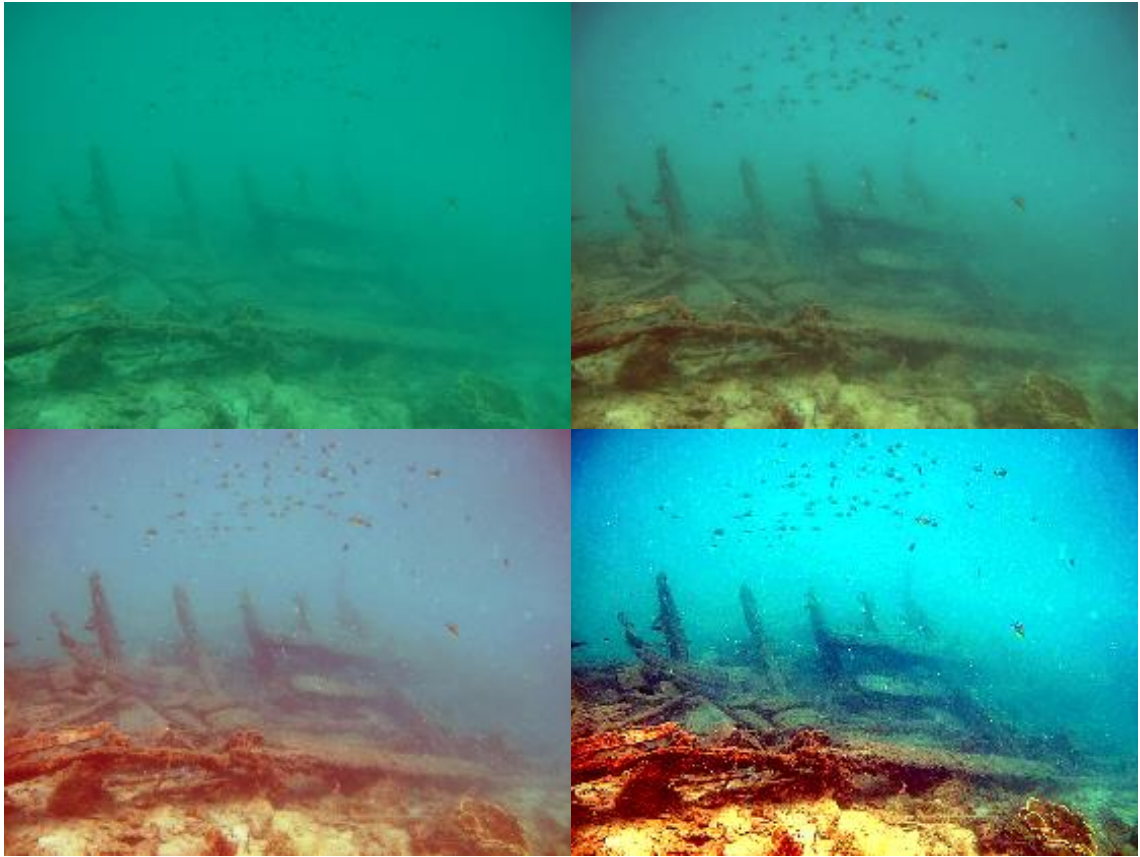

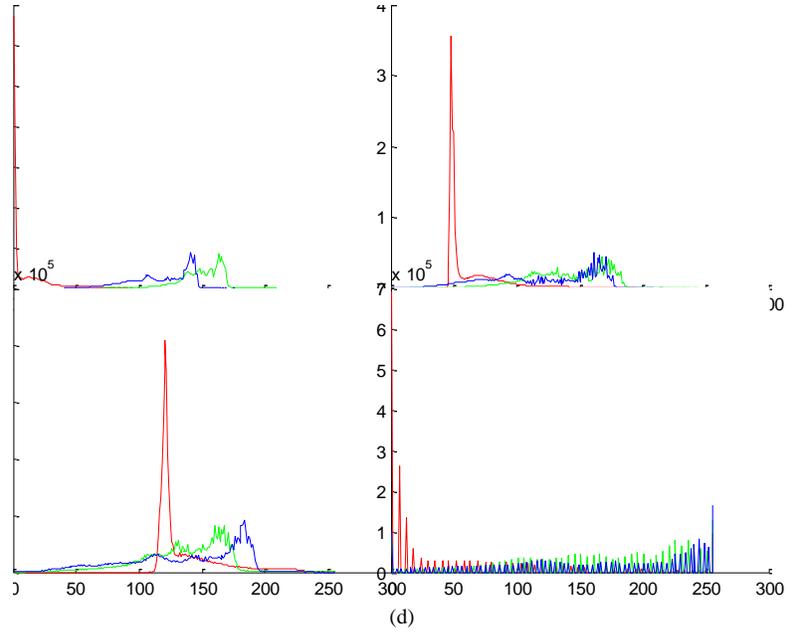

(d)

**KEY**

| Original image | Galdran et al |
|---|---|
| Ancuti et al | PA |

Fig. 2 (d) processed images using various algorithms and key to figures

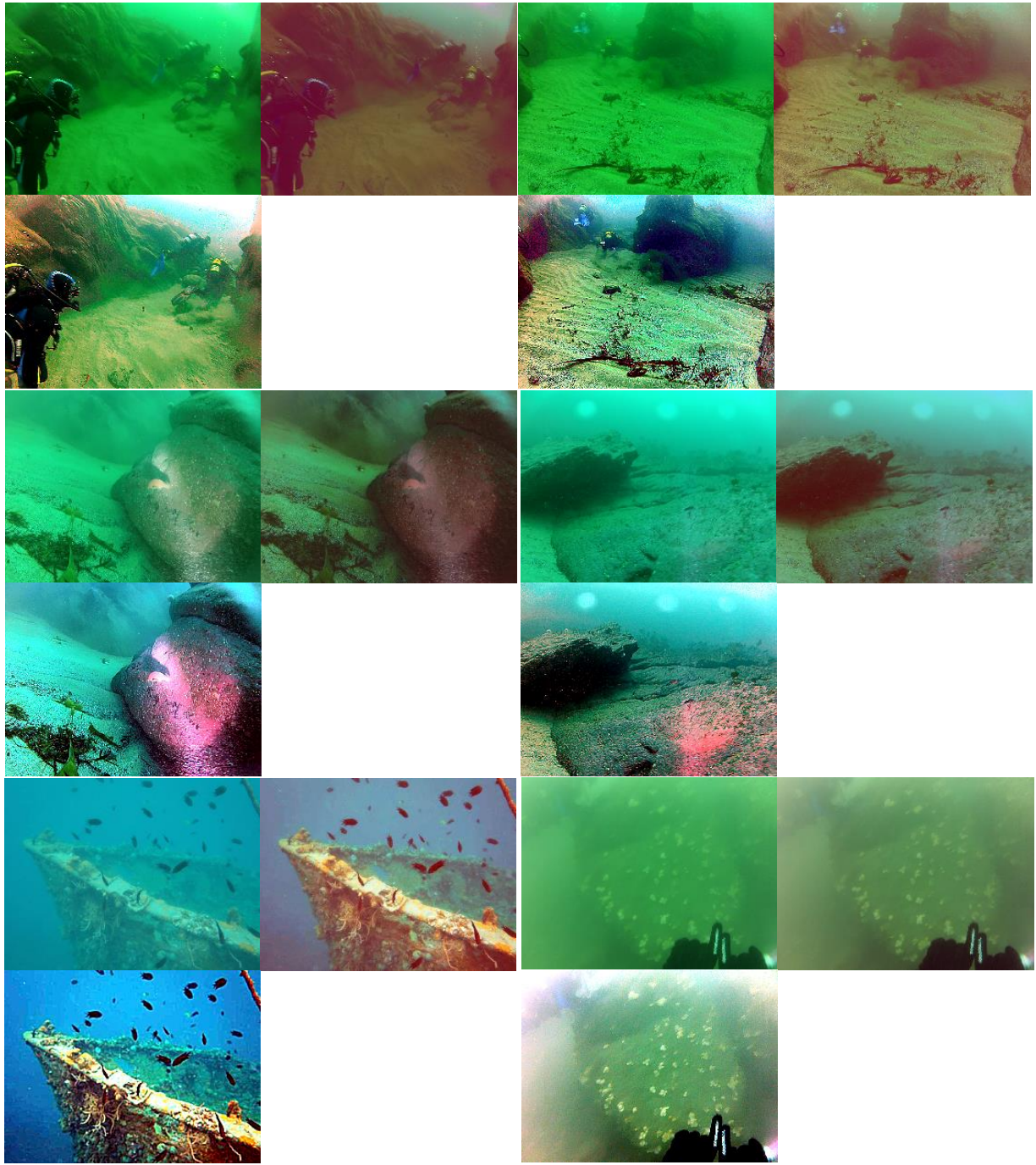

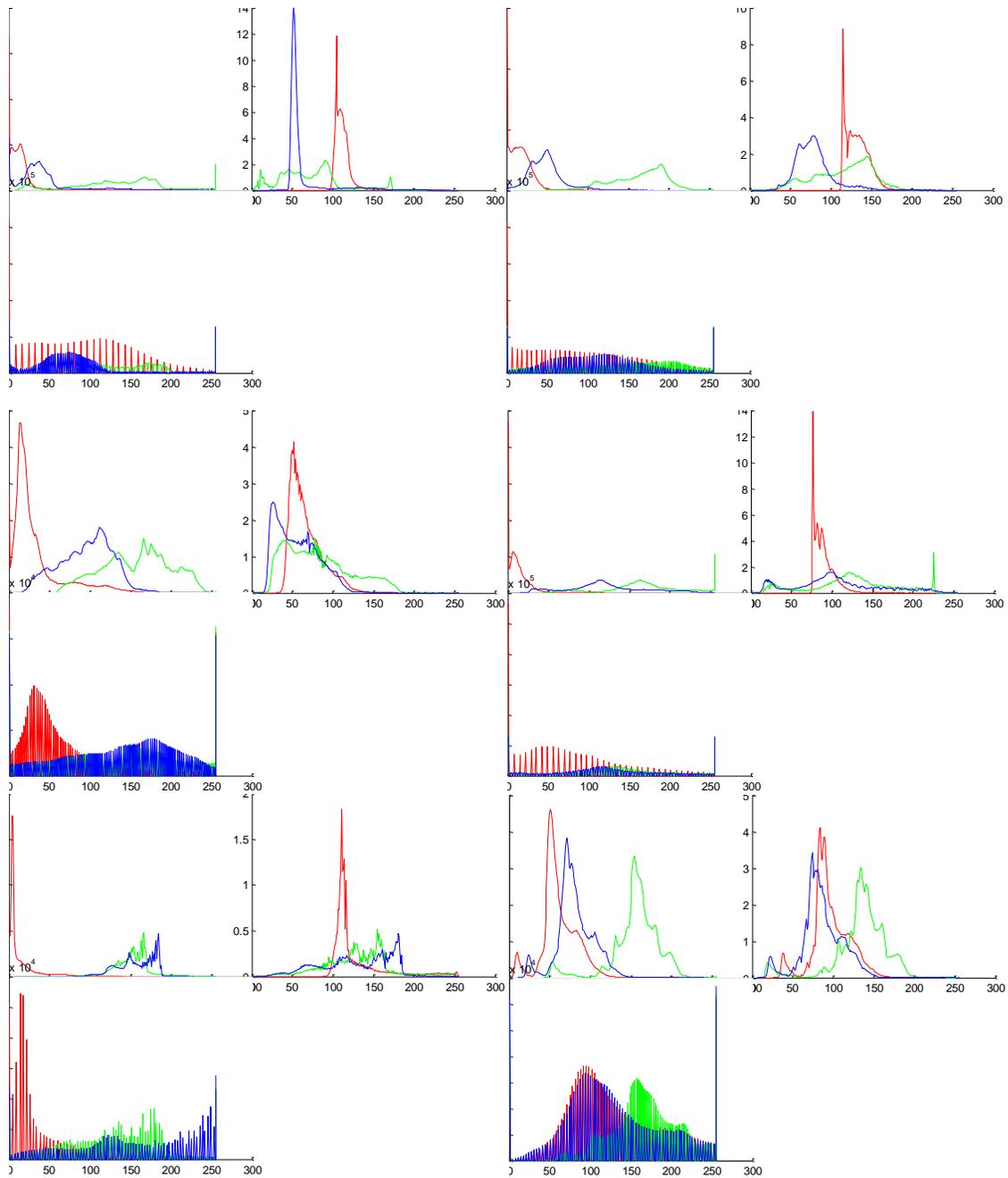

(e)

**KEY**

| Original image | Galdran et al | Original image | Galdran et al |
|---|---|---|---|
| PA | | PA | |
| Original image | Galdran et al | Original image | Galdran et al |
| PA | | PA | |
| Original image | Galdran et al | Original image | Galdran et al |
| PA | | PA | |

Fig. 2 (e) processed images using various algorithms and key to figures

**4.2. Hazy images**

We present results for sample hazy images processed with PA-2 in this section and compare the proposed approach to several existing algorithms from the literature. The algorithms compared include those by Dong et al [20], Ancuti et al [8], Kratz and Nishino [21], Zhang et al [22], Oakley and Bu [23], Kim et al [24], Hsieh et al [25], Meng et al [26], Gibson and Nguyen [27], Yang et al [28], Guo et al [29], Anwar and Khosla [30], Liu et al [31], Ju et al [32], Kopf et al [33], Tan [34], Fattal [35], Tarel and Hautiere [36], Zhu, et al [37], He, et al [38], Ren, et al [3], Dai & Tarel [39], Nishino, et al [40], Galdran, et al [41], Wang and He [42], AMEF [43], PDE-GOC-SSR-CLAHE/PDE-Retinex [44], PDE-IRCES [45], FMIRCES [46] [47] against PA-2 (with $k = 1$ in this case to yield balanced results).

In Fig. 2, PA-1 shows considerable colour correction and local contrast enhancement compared to most of the listed algorithms. This is in spite of its relatively lower complexity compared to the deep neural network-based approaches. This is remarkable since several of these algorithms are highly computationally and structurally complex. With no available runtime information for these particular set of algorithms, it is not possible to compare their execution times with PA-1.

For the *Tiananmen* image in Fig. 3, best results are observed with Zhu, et al, He, et al, Ren, et al, PA-2, PDE-GOC-SSR-CLAHE/PDE-Retinex, which exhibit improved contrast and detail enhancement with no (or minimal) over-enhancement/discolouration of the sky region. The method by He, et al (darker with more halos), and PDE-GOC-SSR-CLAHE both depict visual halos. The PDE-IRCES clearly shows less distortion of sky region but also less contrast in the formerly hazy regions of the image. Tarel and Hautiere show more enhanced details but with discolouration of sky regions. The AMEF once more yields an image with dull and distorted colours.

For the *Toys* image in Fig. 4, the best results are observed with PA-2, PDE-GOC-SSR-CLAHE/PDE-Retinex, algorithms by He et al (visible halos), Wang, et al, Ren et al, Dai, et al and Zhu, et al. The rest depict faded image results and colour distortion with marginal detail enhancement, while AMEF yields a pale image with distorted colours.

In Fig. 5 (*Canyon* image), the best results are those of Ren et al, Zhu et al, PDE-Retinex, FMIRCES, PDE-IRCES in terms of colour and contrast improvement. The others such as Fattal, Tan and Nishino et al result in over-enhancement.

For the *pumpkin*s image in Fig. 6, most methods yield reasonable results. The best are those by Ren et al, Dong et al, Fattal, He et al, PDE-Retinex, Zhu et al, Yeh et al, Ancuti et al, Kratz and Nishino, PDE-IRCES, FMIRCES and PA-2.

In Fig. 7 (*Canon* image), best results include He et al, Nishino et al, Wang et al, PDE-Retinex, FMIRCES, PA-2, Dong et al, PDE-IRCES (though too dark). The results of Anwar and Khosla (colour distortion and over-brightness), Tarel and Hautiere, Galdran et al and Zhu et al (under-enhancement with haze).

In Fig. 8 (*brickhouse* image), generally acceptable results are observed. The best include Zhu et al, He et al, Tarel and Hautiere, Zhang et al, Dong et al, PDE-Retinex, PDE-IRCES, FMIRCES, PA-2 and Hsieh et al. Poor results are observed in Kim et al, Ancuti et al, Yeh et al, Fattal, Ren et al, (which are under-enhanced) Guo et al (over-enhanced) and Liu et al (darkened image details).

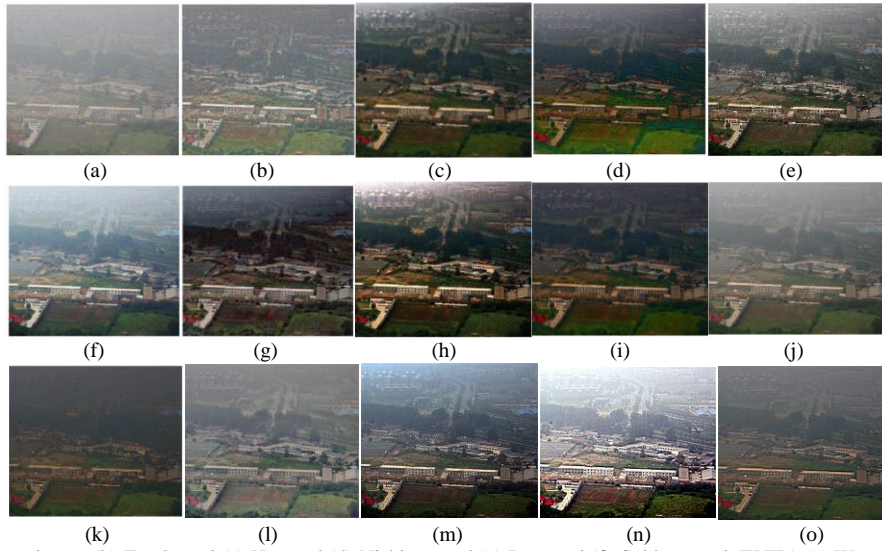

Fig. 3. (a) Original hazy image (b) Tarel, et al (c) He, et al (d) Nishino, et al (e) Ren et al (f) Galdran et al (EVID) (g) Wang & He (h) Dong et al (i) PDE-Retinex (j) Zhu et al (k) PDE-IRCES (l) AMEF (m) FMIRCES (n) PA-2 (o) PA-2 (without GOCS)

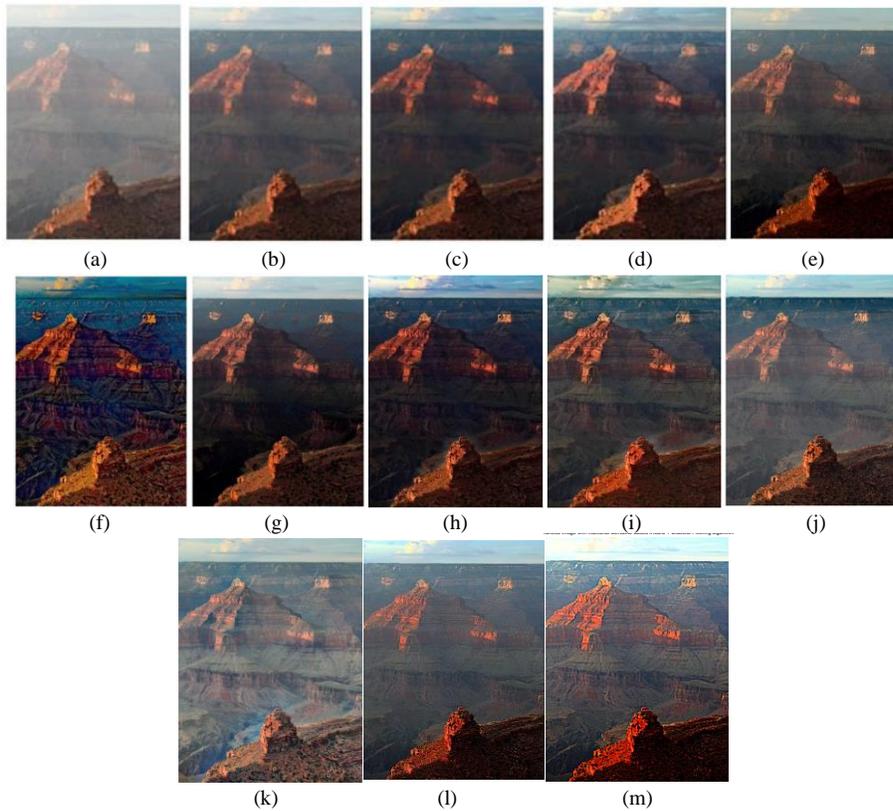

Fig. 4. (a) Original hazy image (b) Fattal (c) Tan (d) Yang et al (e) PDE-IRCES (f) Nishino et al (g) Zhu et al (h) He et al (i) PDE-Retinex (j) Ren et al (k) AMEF (l) FMIRCES (m) PA-2

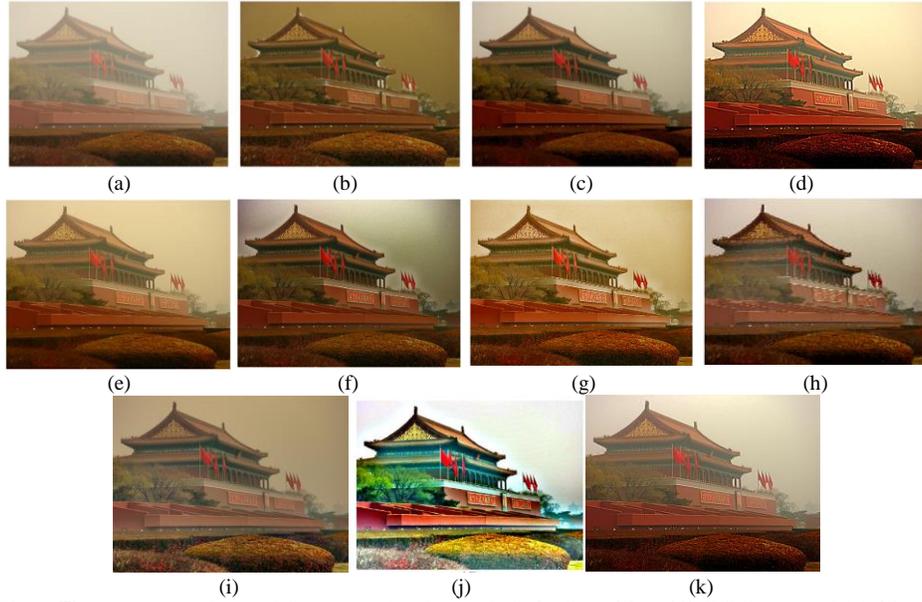

Fig. 5. (a) Original hazy *Tianamen* image (b) Tarel & Hautiere (c) Zhu, et al (d) PA-2 (e) PDE-IRCES-2 (f) He, et al (g) PDE-GOC-SSR-CLAHE (h) Ren et al (i) Galdran (AMEF) (j) Ju et al [32] (k) FMIRCES [46] [47]

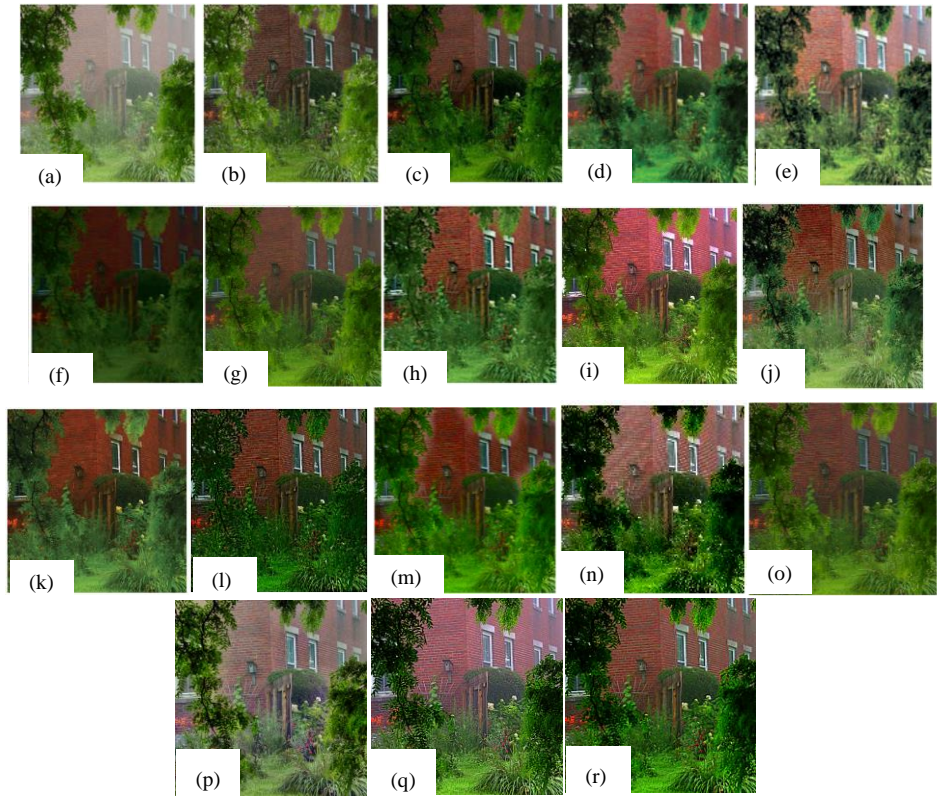

Fig. 6. (a) Original hazy image (b) Hsieh et al (c) Zhu, et al (d) Kim et al (e) Ancuti, et al (f) He et al (g) Tarel, et al (h) Zhang, et al (i) Yeh, et al (j) Fattal (k) Dong, et al (l) Guo, et al (m) PDE-Retinex (n) Ren et al (o) PDE-IRCES (p) AMEF (q) FMIRCES (r) PA-2 (without GOCS)

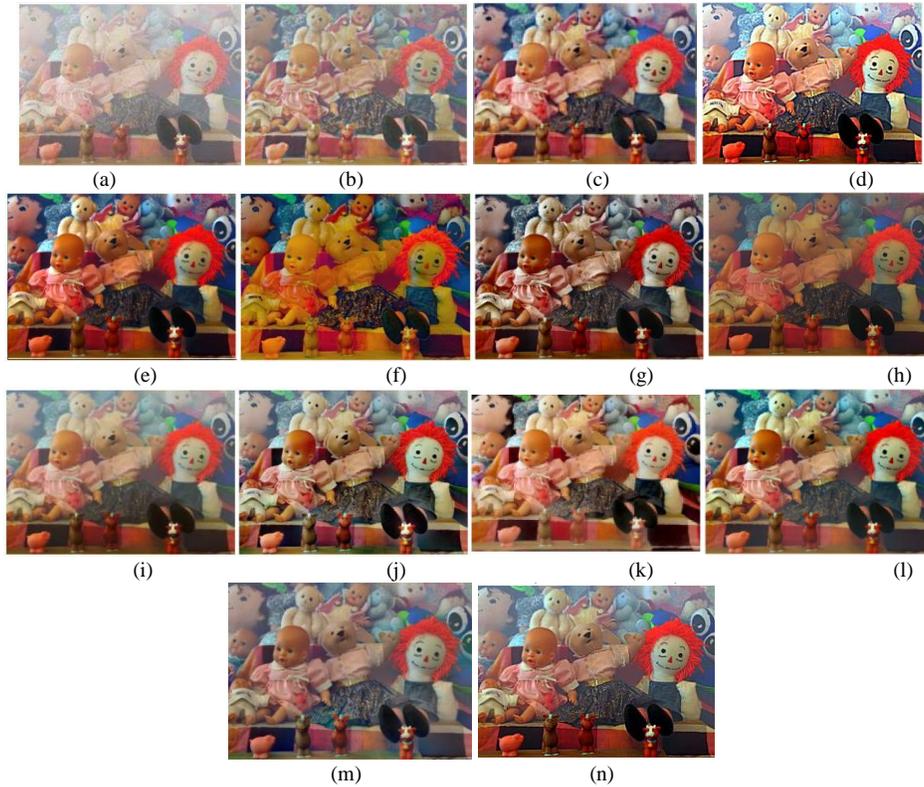

Fig. 7. (a) Original hazy *Toys* image (b) Tarel & Hautiere [36] (c) Dai et al [39] (d) PA-2 (e) He et al [38] (f) Nishino, et al [40] (g) PDE-GOC-SSR-CLAHE [44] (h) PDE-IRCES [45] (i) Galdran, et al (EVID) [41] (j) Wang & He [42] (k) Zhu, et al [37] (l) Ren, et al [3] (m) Galdran (AMEF) [43] (n) FMIRCES [46] [47]

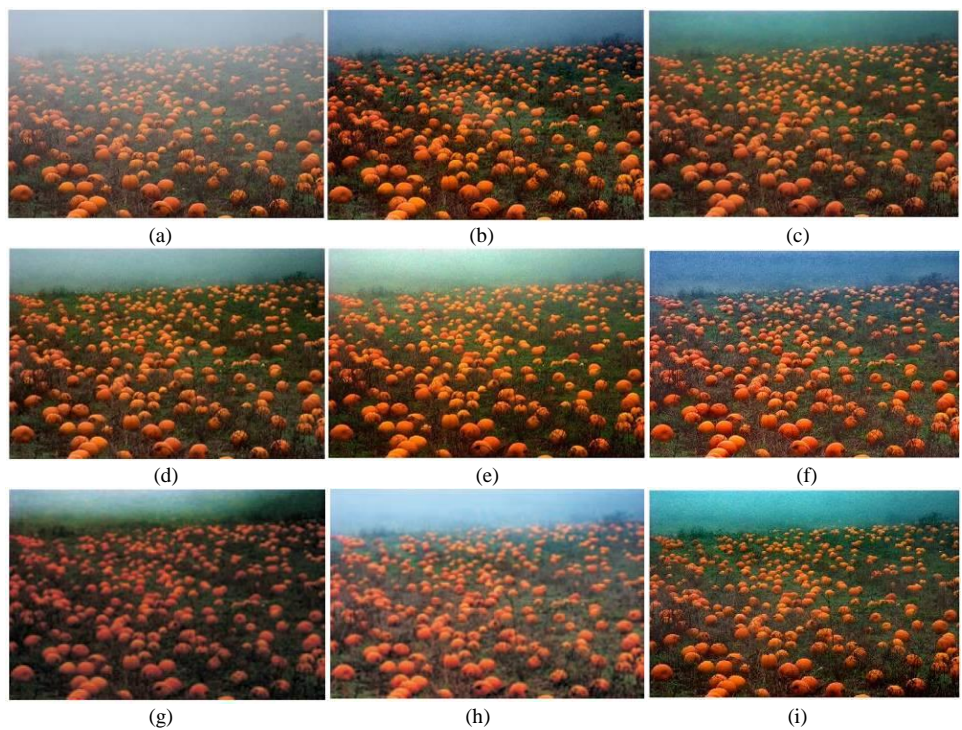

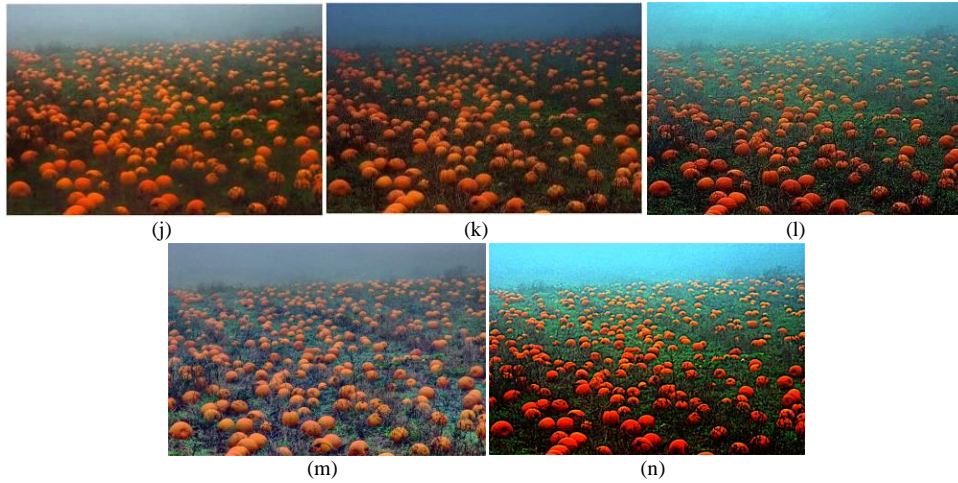

Fig. 8. (a) Original hazy image (b) Fattal (c) Dong et al (d) He et al (e) Yeh et al (f) PDE-Retinex (g) Kratz and Nishino (h) Ancuti et al (i) Ren et al (j) Zhu et al (k) PDE-IRCES (l) FMIRCES (m) AMEF (n) PA-2

In Fig. 9, we show the colour histograms for 99 hazy images processed with FMIRCES and PA and the same trend observed in enhanced underwater images is also seen here. The histograms of images processed with PA are highly stretched compared to images processed with FMIRCES. Furthermore, values for images processed with PA are much more spread out indicating increased standard deviation, variance and contrast. Thus, the improvements of PA over FMIRCES is also observed in the histogram plot comparison.

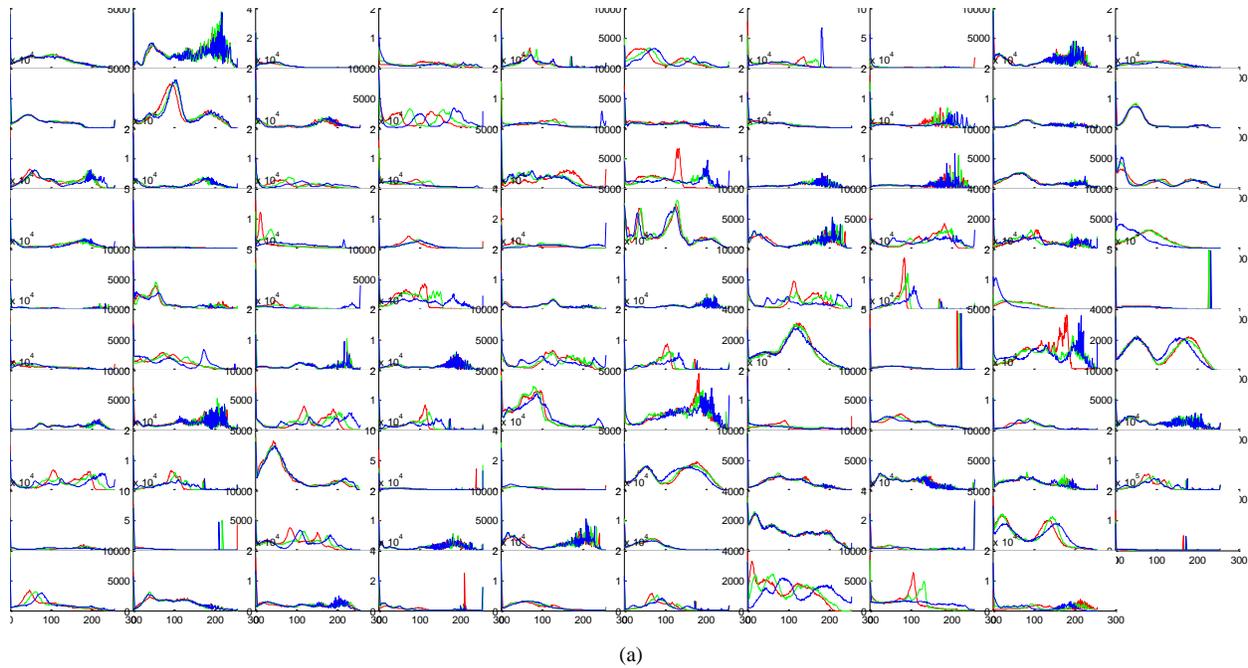

(a)

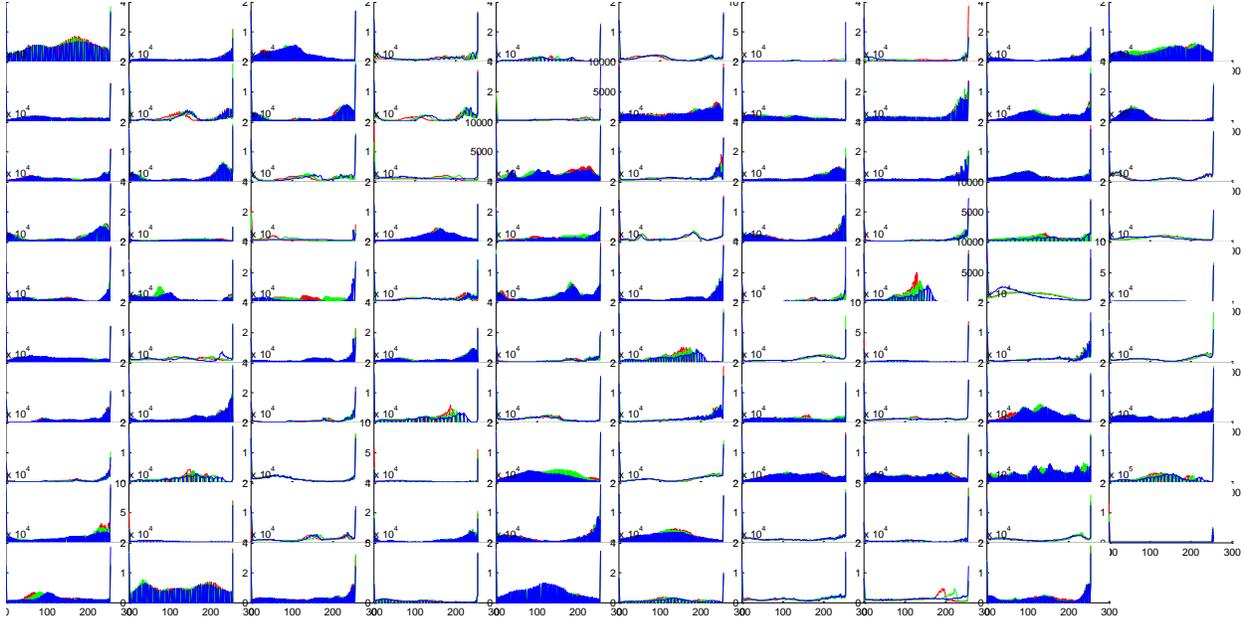
(b)
Fig. 9. Image colour histograms of hazy images processed with (a) FMIRCES and (b) PA

In Table 3, we compare the average runtimes of the various available de-hazing algorithms with PA. Results indicate that PA has the second fastest runtime compared to the other algorithms. It also yields better results than the fastest algorithm from previous work (which is the FMIRCES) [46] by sacrificing some speed for improved localized contrast and better colour rendition.

Table 3. Average runtimes (seconds) for images processed with available algorithm implementations

| Algos | He, et al | Zhu, et al | Ren et al | AMEF | PDE-GOC-SSR-CLAHE | PDE-IRCES | FMIRCES | PA |
|---|---|---|---|---|---|---|---|---|
| Mean times(s) | 1.230375 | 0.959301 | 2.500416 | 1.4427 | 3.544029 | 2.321865 | **0.421316** | **0.832497** |

## 5. Conclusion

Based on the results, we can see that the proposed algorithm improves on the initial formulation by addressing the local contrast and edge enhancement in addition to image brightening. The image histogram and statistics indicate improvements in addition to the objective metrics used to corroborate the results. The proposed approach further validates the use of tonal correction and mapping operators as viable alternatives to the image de-hazing problem. Moreover, the proposed approaches surpass a majority of contemporary and much more complex underwater and hazy image enhancement algorithms from the literature. The proposed schemes are versatile since they process both hazy and underwater images adequately. These outcomes are verified in terms of contrast and colour enhancement/correction, with good visual and objective results and minimized runtime.